\documentclass[sn-mathphys,Numbered]{sn-jnl}
\UseRawInputEncoding
\usepackage{graphicx}%
\usepackage{multirow}%
\usepackage{amsmath,amssymb,amsfonts}%
\usepackage[title]{appendix}%
\usepackage{xcolor}%
\usepackage{textcomp}%
\usepackage{manyfoot}%
\usepackage{listings}%

\begin{document}

\title[Article Title]{CCAT: Comparisons of 280 GHz TiN and Al Kinetic Inductance Detector Arrays}


\author*[1]{\fnm{Cody J.} \sur{Duell}}\email{cjd259@cornell.edu}
\author[2]{\fnm{Jason} \sur{Austermann}}
\author[2]{\fnm{James} \sur{Beall}}
\author[3]{\fnm{James R.} \sur{Burgoyne}}
\author[4]{\fnm{Scott C.} \sur{Chapman}}
\author[1, 5]{\fnm{Steve K.} \sur{Choi}}
\author[5]{\fnm{Rodrigo G.} \sur{Freundt}}
\author[6]{\fnm{Jiansong} \sur{Gao}}
\author[7]{\fnm{Christopher} \sur{Groppi}}
\author[8]{\fnm{Anthony I.} \sur{Huber}}
\author[1]{\fnm{Zachary B.} \sur{Huber}}
\author[2]{\fnm{Johannes} \sur{Hubmayr}}
\author[1]{\fnm{Ben} \sur{Keller}}
\author[1]{\fnm{Yaqiong} \sur{Li}}
\author[1]{\fnm{Lawrence T.} \sur{Lin}}
\author[7]{\fnm{Justin} \sur{Matthewson}}
\author[7]{\fnm{Philip} \sur{Mauskopf}}
\author[1]{\fnm{Alicia} \sur{Middleton}}
\author[1]{\fnm{Colin C.} \sur{Murphy}}
\author[1, 5]{\fnm{Michael D.} \sur{Niemack}}
\author[9]{\fnm{Thomas} \sur{Nikola}}
\author[3]{\fnm{Adrian K.} \sur{Sinclair}}
\author[1]{\fnm{Ema} \sur{Smith}}
\author[2]{\fnm{Jeff} \sur{van Lanen}}
\author[2]{\fnm{Anna} \sur{Vaskuri}}
\author[1]{\fnm{Eve M.} \sur{Vavagiakis}}
\author[2]{\fnm{Michael} \sur{Vissers}}
\author[1]{\fnm{Samantha} \sur{Walker}}
\author[2]{\fnm{Jordan} \sur{Wheeler}}
\author[10]{\fnm{Bugao} \sur{Zou}}

\affil[1]{\orgdiv{Department of Physics}, \orgname{Cornell University}, \orgaddress{\city{Ithaca}, \state{NY}, \country{USA}}}
\affil[2]{\orgdiv{Quantum Sensors Division}, \orgname{NIST}, \orgaddress{\city{Boulder}, \state{CO}, \country{USA}}}
\affil[3]{\orgdiv{Department of Physics and Astronomy}, \orgname{University of British Columbia}, \orgaddress{\city{Vancouver},\state{BC}, \country{Canada}}}
\affil[4]{\orgdiv{Department of Physics and Atmospheric Science}, \orgname{Dalhousie University}, \orgaddress{\city{Halifax},\state{NS}, \country{Canada}}}
\affil[5]{\orgdiv{Department of Astronomy}, \orgname{Cornell University}, \orgaddress{\city{Ithaca},\state{NY}, \country{USA}}}
\affil[6]{\orgdiv{AWS Center for Quantum Computing}, \orgname{Amazon}, \orgaddress{\city{Pasadena}, \state{CA}, \country{USA}}}
\affil[7]{\orgdiv{School of Earth and Space Exploration}, \orgname{Arizona State University}, \orgaddress{\city{Tempe}, \state{AZ}, \country{USA}}}
\affil[8]{\orgdiv{Department of Physics and Astronomy}, \orgname{University of Victoria}, \orgaddress{\city{Victoria}, \state{BC}, \country{Canada}}}
\affil[9]{\orgdiv{Cornell Center for Astrophysics and Planetary Sciences}, \orgname{Cornell University}, \orgaddress{\city{Ithaca}, \state{NY}, \country{USA}}}
\affil[10]{\orgdiv{Department of Applied and Engineering Physics}, \orgname{Cornell University}, \orgaddress{\city{Ithaca}, \state{NY}, \country{USA}}}

\abstract{The CCAT Collaboration's six-meter Fred Young Submillimeter Telescope is scheduled to begin observing in the Chilean Atacama in 2025, targeting a variety of science goals throughout cosmic history. Prime-Cam is a 1.8-meter diameter cryostat that will host up to seven independent instrument modules designed for simultaneous spectroscopic and broadband, polarimetric surveys at millimeter to submillimeter wavelengths. The first of these instrument modules, the 280 GHz module, will include ${\sim}$10,000 kinetic inductance detectors (KIDs) across three arrays. While the first array was fabricated out of tri-layer TiN/Ti/TiN, the other two arrays were fabricated out of a single layer of Al. This combination of materials within the same instrument provides a unique opportunity to directly compare the performance and noise properties of two different detector materials that are seeing increasing use within the field. We present preliminary comparisons here based on lab testing, along with a discussion of the potential impacts on operation when observing and translating raw data to science-grade maps.}



\maketitle
\section{Introduction}\label{sec1}

Superconducting detectors have become the state of the art for measuring faint astronomical signals at millimeter and submillimeter wavelengths. Kinetic inductance detectors (KIDs), a type of superconducting resonator \cite{day_broadband_2003}, have become an increasingly popular choice in recent years due to their sensitivity and ease of multiplexing. KIDs allow for high detector counts and photon-limited sensitivity, while also being simpler to read out than comparably-sensitive transition edge sensors. At the same time, the technology is new enough that there remain significant questions about their noise performance and optimization for the field, one of the most basic of which is the choice of material for the detector. For photon-limited operation of KIDs at millimeter and submillimeter wavelengths, KIDs have typically been designed for bath temperatures of 100--300 mK with an inductor transition temperature, $T_c$, of 0.5--1.5 K \cite{dober_next-generation_2014, adam_nika2_2018, austermann_millimeter-wave_2018, wheeler_broadband_2022}. At the same time, the material must also be robust to fabrication at wafer-scales and repeated cryogenic cycling.

Given these constraints, two of the most popular materials for KIDs in these bands are Al and TiN, both of which have shown photon-limited noise performance \cite{mauskopf_photon-noise_2014, hubmayr_photon-noise_2015} and have been fabricated at array scales. The primary difference between these materials is that TiN is a “disordered” superconductor, meaning that it has much higher resistivity above $T_c$ than Al, which has several implications, including higher kinetic inductance per square and easier impedance matching for photon absorption. Imaging experiments that have fielded Al KIDs include NIKA \cite{monfardini_dual-band_2011}, NIKA2 \cite{adam_nika2_2018}, OLIMPO \cite{paiella_kinetic_2019}, and MUSCAT \cite{brien_muscat_2018}, while TiN detectors have seen use for MAKO \cite{swenson_mako_2012}, BLAST-TNG \cite{dober_next-generation_2014}, TolTEC \cite{wilson_toltec_2020}, and, at slightly higher frequencies, ARCONS \cite{mazin_arcons_2013}. 

CCAT's 280 GHz instrument module for Prime-Cam provides an opportunity to test these two materials side-by-side in a near 1:1 setting. In this paper, we will lay the groundwork for that comparison, as well as provide early discussion based on lab testing. In Section 2, we will briefly describe Prime-Cam and the 280 GHz instrument module, as well as the relevant focal plane and detector designs. In Section 3, we will make early comparisons in terms of noise performance, sensitivity, and expected yield. We will conclude with a discussion of future work. 

\begin{figure}[ht]%
\centering
\includegraphics[width=0.9\textwidth]{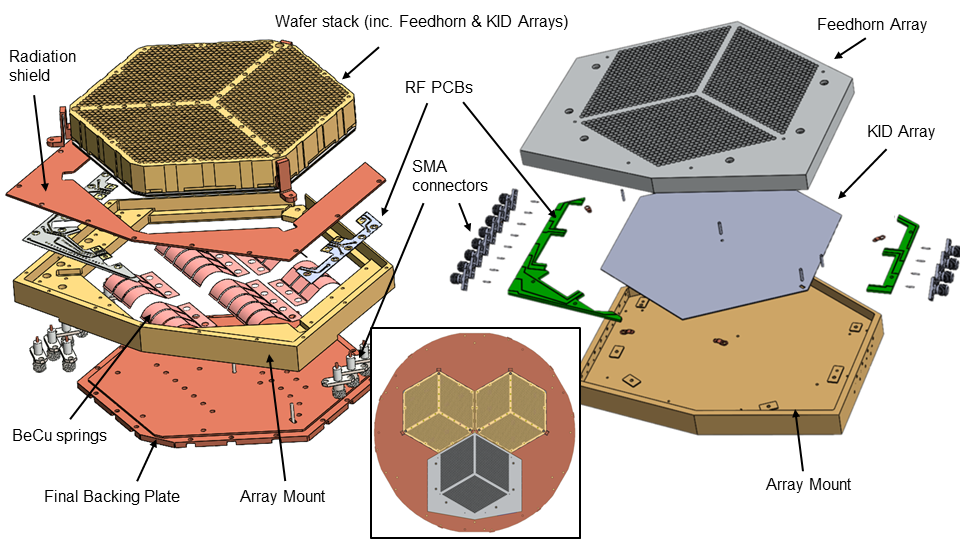}
\caption{Exploded views of the Al detector modules (left) with Si platelet feedhorns and TiN detector  module (right) with Al machined feedhorns. The inset shows the focal plane layout for all three arrays.}\label{fig1}
\end{figure}

\section{Background}\label{sec2}

\subsection{Prime-Cam and the Fred Young Submillimeter Telescope}\label{subsec1}

The Fred Young Submillimeter Telescope (FYST) is being built in the Chilean Atacama by  the CCAT Collaboration at an elevation of 5600 meters near the summit of Cerro Chajnantor \cite{parshley_ccat-prime_2018}. It has a six-meter aperture with a modified off-axis, crossed-Dragone design \cite{niemack_designs_2016}, which offers a wide field-of-view and high throughput to take advantage of the exceptional atmospheric conditions. FYST is scheduled to begin making observations in 2025. Prime-Cam will be one of two primary instruments for FYST, targeting a variety of science goals throughout cosmic history \cite{vavagiakis_prime-cam_2018, collaboration_ccat-prime_2022}. Prime-Cam is a 1.8-m diameter cryostat that will host up to seven independent instrument modules spanning 220--850 GHz, each with a ${\sim}1.3$ degree field-of-view, allowing for the completion of simultaneous broadband and spectroscopic surveys \cite{choi_sensitivity_2020}. 

\subsection{280 GHz Instrument Module and Mod-Cam}\label{subsec2}

The first of the Prime-Cam instrument modules, the 280 GHz module, will be used for both wide-field and small-field surveys and is populated with three independent, hexagonally-tiled detector arrays with a shared optical path. The 280 GHz instrument module will include ${\sim}10,000$ polarization-sensitive KIDs across three arrays and will initially be deployed in Mod-Cam, a single module testbed for telescope commissioning and development of future modules. Details of Mod-Cam and the 280 GHz instrument module can be found in \cite{vavagiakis_ccat-prime_2022}.  

\subsection{Focal Plane Module Designs}\label{subsec3}

The first KID array was fabricated out of tri-layer TiN/Ti/TiN, while the other two arrays were entirely out of Al. All three arrays were fabricated on 550-micron silicon-on-insulator wafers by the Quantum Sensors Group at the National Institute for Standards and Technology (NIST) in Boulder, CO. While the same optics are used between the two types of arrays, there are some differences in the focal plane module designs. Both package designs are shown in Figure 1 in an exploded view. The TiN array utilizes a gold-plated aluminum package and machined Al feedhorns. Optical alignment is set by a combination of pin-and-slot design and careful machining, while vibrational variation in the waveguide gaps above the detectors is suppressed by using pogo-pins from the bottom of the feedhorn array to hold the array against the module package. More details can be found in \cite{duell_ccat-prime_2020}. 

In comparison, both Al arrays use gold-plated Si-platelet feedhorns with gold-plated copper packaging. Optical alignment is set while warm by use of alignment pins, verified visually under a microscope, and maintained through cooling due to the matched coefficients of thermal contraction. The waveguide gap is set by an interface plate and  the detector array, feedhorn chokes, interface wafer, and backing plate are held against the back of the Si feedhorn array by flexible springs, in the form of BeCu fingerstock. This design shares many elements in common with the array packaging used by TolTEC, as described in \cite{austermann_toltec_2020}.

Both feedhorns are based on the same numerically-optimized spline profile used on TolTEC’s 1.1 mm array \cite{simon_feedhorn_2018, austermann_millimeter-wave_2018}, and the differences in performance are expected to be minimal based on simulations and early measurements. As suggested above, the different choice of feedhorn materials requires unique mechanical alignment and assembly methods between the two array types. Lastly, the interface PCB design was slightly modified for the Al arrays.

\begin{figure}[ht]%
\centering
\includegraphics[width=0.9\textwidth]{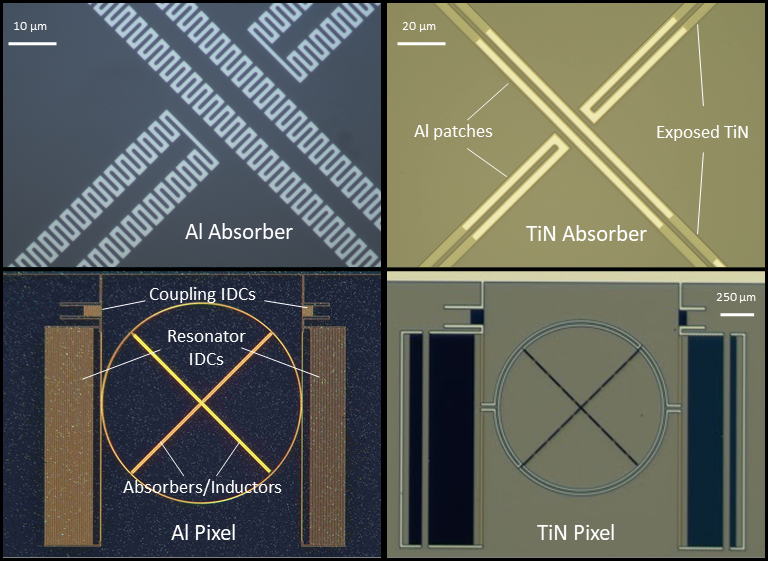}
\caption{TiN pixel design (bottom right) showing two polarization-sensitive detectors. (Bottom left) Al pixel design shown at a similar scale. Close-ups of the respective absorbers are shown on the top, displaying both the meandered Al absorber and the TiN absorbers with Al patches.}\label{fig2}
\end{figure}

\subsection{Pixel Designs}\label{subsec4}

Both pixel types are lumped element polarimeters evolved from previous NIST designs. A single pixel contains two detectors with front-side illuminated, orthogonal antennas that are impedance-matched to the feedhorn-coupled waveguide. The primary components include a polarization-sensitive antenna that serves as both the inductor and a direct absorber, an interdigitated capacitor (IDC) for coupling to the readout line, and an IDC to set the total capacitance. On the backside, the wafer is deep-etched down to the 80-$\mu m$ device layer and deposited with an aluminum ground plane, making a quarter-wavelength reflective backshort to improve the optical coupling. Sample devices are shown in Figure 2. 

The Al devices are passivated with a layer of amorphous silicon and have a $T_c\sim1.4$ K and sheet resistance of $R_s \sim 1$ $\Omega/\square$ (Ohms per square). The TiN/Ti/TiN trilayer devices have a $T_c \sim 1.1$ K, and $R_s \sim 90$ $\Omega/\square$. Due to Al having much lower normal resistance, the Al inductor is meandered with finer traces (${\sim} 1$ $\mu m$ vs. ${\sim} 4$ $\mu m$ linewidth) (as shown in Figure 2) to better match the waveguide impedance. To tune the desired absorber volume while balancing responsivity and coupling efficiency, 100-nm thick patches of Al are placed above the TiN multilayer absorber to act as a short without significantly affecting the impedance as done in the TolTEC designs \cite{austermann_millimeter-wave_2018}. Finally, the Al detectors have slightly larger capacitor geometries. Additional details can be found in \cite{austermann_millimeter-wave_2018, choi_ccat-prime_2022, austermann_aluminum-based_2022}.

\section{Discussion}\label{sec3}

Lab testing of individual witness pixels and full arrays is ongoing at NIST and Cornell, and both detector designs are still planned for deployment. Yields for both types of arrays are expected to be 95\% or better following LED mapping and post-fabrication editing \cite{liu_superconducting_2017}, which is currently in progress for all three arrays. Measured quality factors, responsivity, and noise have met expectations \cite{collaboration_ccat-prime_2022}, though it is worth noting that the Al detectors have slightly higher internal quality factors under load (significantly higher when dark). While both detectors are expected to return science-grade data, there are several differences in performance that can be expected to impact operation or final map-making.

\begin{figure} [ht]
\centering
\includegraphics[width=0.95\textwidth]{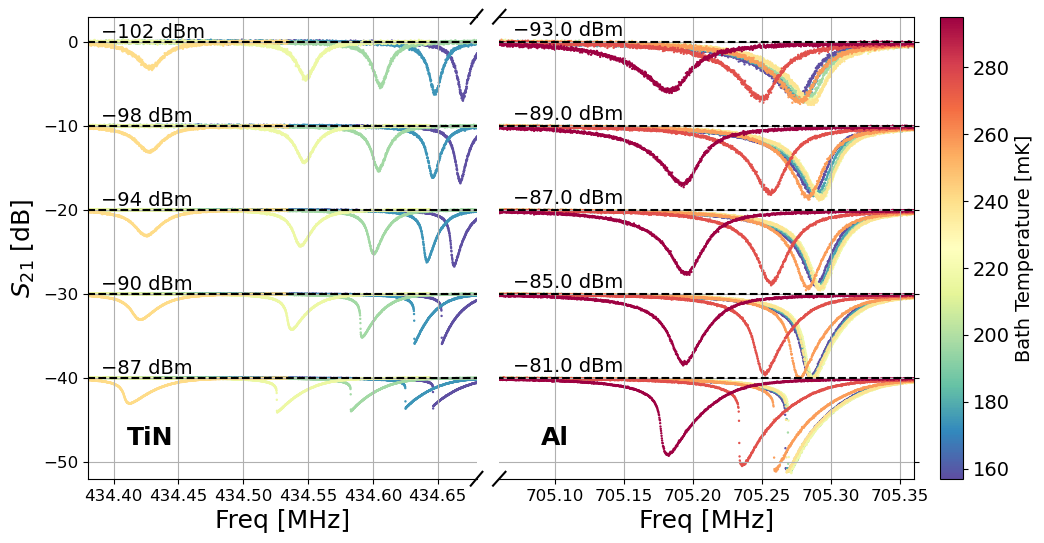}
\label{fig3} 
\caption{ Typical response to changing tone power and bath temperature for TiN (left) and Al (right) with a 10 dB offset between powers for readability. The response of TiN is well fit accounting for nonlinear kinetic inductance \cite{swenson_operation_2013}, leading to a monotonic frequency shift with both tone power and temperature, while total Q decreases with increasing temperature and remains constant with tone power. In contrast, the Al resonant frequency increases between $-93$ dBm and $-87$ dBm, before decreasing at higher powers up to bifurcation, while the Q continually increases with tone power. Similarly, the Q and frequency increase with bath temperature up to ${\sim} 220$ mK at all powers. Tone power is estimated at the detector and should be accurate within 3 dB.}
\end{figure}

The noise profile of the two materials is distinct, particularly at low frequencies, which is one of the features that motivated a shift towards Al detectors. The Al detectors have been measured to have exceptionally low 1/f noise \cite{austermann_aluminum-based_2022}, whereas the TiN have a higher photon-activated noise contribution at low frequencies. This is particularly relevant for wide-field surveys and reconstructing large angular scales, as for science cases involving the cosmic microwave background.

Additionally, the different optical response curves of the materials has been well documented \cite{hubmayr_photon-noise_2015,flanigan_photon_2016}. While TiN KIDs show an unexpectedly linear response to optical illumination, Al detectors have a square-root dependence. This means detector tuning requirements during observations may differ between arrays.

One final point of discussion is the tone power sensitivity and nonlinearity. Optimizing tone power is an integral part of achieving the highest signal-to-noise for KIDs in the field. It is generally preferable to read out KIDs at the highest tone powers attainable without reaching bifurcation, to both maximize signal-to-noise going into the first-stage amplifier and minimize TLS noise. On this count, it is useful that the Al detectors bifurcate at higher tone powers. However, as a result of the interplay between the nonlinear kinetic inductance \cite{swenson_operation_2013} and nonequilibrium quasiparticle dynamics \cite{de_visser_evidence_2014}, the Al detectors show more complicated interactions between tone power linearity, optical loading, and the observed resonator parameters, requiring careful tuning. Some of this is seen in Figure 3, which shows the non-monotonic relationship between tone-power and measured resonant frequency and Q. This also significantly distorts the resonator line shape, making it difficult to fit for these parameters. This sensitivity to tone power will necessitate frequent tuning as observing conditions vary, or resonator tone-tracking, which is planned for implementation after deployment \cite{sinclair_ccat-prime_2022}. How this will impact observing efficiency and on-sky noise performance remains to be seen.

\section{Conclusion}

The 280 GHz instrument module in Prime-Cam will be an exceptional opportunity to directly compare the performance and practical operation of similarly optimized Al and TiN KIDs. We have described the background of the detector designs and several of the relevant features that are expected to differentiate their performance. All three of the arrays have been fabricated and are currently undergoing LED-mapping and post-fabrication editing to achieve targeted on-sky yield \cite{liu_superconducting_2017}. Once completed, we will be able to fully optically characterize the arrays, including measurements of polarization efficiency, cross-polarization, responsivity, and noise prior to deployment in Mod-Cam on FYST for commissioning.

\backmatter

\bmhead{Acknowledgments}

The CCAT project, FYST and Prime-Cam instrument have been supported by generous contributions from the Fred M. Young, Jr. Charitable Trust, Cornell University, and the Canada Foundation for Innovation and the Provinces of Ontario, Alberta, and British Columbia. The construction of the FYST telescope was supported by the Gro{\ss}ger{\"a}te-Programm of the German Science Foundation (Deutsche Forschungsgemeinschaft, DFG) under grant INST 216/733-1 FUGG, as well as funding from Universit{\"a}t zu K{\"o}ln, Universit{\"a}t Bonn and the Max Planck Institut f{\"u}r Astrophysik, Garching. MDN acknowledges support from NSF grant AST-2117631. EMV acknowledges support from NSF award AST-2202237. ZBH was supported by a NASA Space Technology Graduate Research Opportunity. SKC acknowledges support from NSF award AST2001866. SW acknowledges support from the Cornell CURES fellowship.

\bibliography{references.bib}


\begin{thebibliography}{29}
\ifx \bisbn   \undefined \def \bisbn  #1{ISBN #1}\fi
\ifx \binits  \undefined \def \binits#1{#1}\fi
\ifx \bauthor  \undefined \def \bauthor#1{#1}\fi
\ifx \batitle  \undefined \def \batitle#1{#1}\fi
\ifx \bjtitle  \undefined \def \bjtitle#1{#1}\fi
\ifx \bvolume  \undefined \def \bvolume#1{\textbf{#1}}\fi
\ifx \byear  \undefined \def \byear#1{#1}\fi
\ifx \bissue  \undefined \def \bissue#1{#1}\fi
\ifx \bfpage  \undefined \def \bfpage#1{#1}\fi
\ifx \blpage  \undefined \def \blpage #1{#1}\fi
\ifx \burl  \undefined \def \burl#1{\textsf{#1}}\fi
\ifx \doiurl  \undefined \def \doiurl#1{\url{https://doi.org/#1}}\fi
\ifx \betal  \undefined \def \betal{\textit{et al.}}\fi
\ifx \binstitute  \undefined \def \binstitute#1{#1}\fi
\ifx \binstitutionaled  \undefined \def \binstitutionaled#1{#1}\fi
\ifx \bctitle  \undefined \def \bctitle#1{#1}\fi
\ifx \beditor  \undefined \def \beditor#1{#1}\fi
\ifx \bpublisher  \undefined \def \bpublisher#1{#1}\fi
\ifx \bbtitle  \undefined \def \bbtitle#1{#1}\fi
\ifx \bedition  \undefined \def \bedition#1{#1}\fi
\ifx \bseriesno  \undefined \def \bseriesno#1{#1}\fi
\ifx \blocation  \undefined \def \blocation#1{#1}\fi
\ifx \bsertitle  \undefined \def \bsertitle#1{#1}\fi
\ifx \bsnm \undefined \def \bsnm#1{#1}\fi
\ifx \bsuffix \undefined \def \bsuffix#1{#1}\fi
\ifx \bparticle \undefined \def \bparticle#1{#1}\fi
\ifx \barticle \undefined \def \barticle#1{#1}\fi
\bibcommenthead
\ifx \bconfdate \undefined \def \bconfdate #1{#1}\fi
\ifx \botherref \undefined \def \botherref #1{#1}\fi
\ifx \url \undefined \def \url#1{\textsf{#1}}\fi
\ifx \bchapter \undefined \def \bchapter#1{#1}\fi
\ifx \bbook \undefined \def \bbook#1{#1}\fi
\ifx \bcomment \undefined \def \bcomment#1{#1}\fi
\ifx \oauthor \undefined \def \oauthor#1{#1}\fi
\ifx \citeauthoryear \undefined \def \citeauthoryear#1{#1}\fi
\ifx \endbibitem  \undefined \def \endbibitem {}\fi
\ifx \bconflocation  \undefined \def \bconflocation#1{#1}\fi
\ifx \arxivurl  \undefined \def \arxivurl#1{\textsf{#1}}\fi
\csname PreBibitemsHook\endcsname

\bibitem[\protect\citeauthoryear{Day et~al.}{}]{day_broadband_2003}
\begin{botherref}
\oauthor{\bsnm{Day}, \binits{P.K.}},
\oauthor{\bsnm{{LeDuc}}, \binits{H.G.}},
\oauthor{\bsnm{Mazin}, \binits{B.A.}},
\oauthor{\bsnm{Vayonakis}, \binits{A.}},
\oauthor{\bsnm{Zmuidzinas}, \binits{J.}}:
A broadband superconducting detector suitable for use in large arrays
\textbf{425}(6960),
817--821
\doiurl{10.1038/nature02037} .
Number: 6960 Publisher: Nature Publishing Group
\end{botherref}
\endbibitem

\bibitem[\protect\citeauthoryear{Dober et~al.}{}]{dober_next-generation_2014}
\begin{botherref}
\oauthor{\bsnm{Dober}, \binits{B.J.}},
\oauthor{\bsnm{Ade}, \binits{P.A.R.}},
\oauthor{\bsnm{Ashton}, \binits{P.}},
\oauthor{\bsnm{Angilè}, \binits{F.E.}},
\oauthor{\bsnm{Beall}, \binits{J.A.}},
\oauthor{\bsnm{Becker}, \binits{D.}},
\oauthor{\bsnm{Bradford}, \binits{K.J.}},
\oauthor{\bsnm{Che}, \binits{G.}},
\oauthor{\bsnm{Cho}, \binits{H.-M.}},
\oauthor{\bsnm{Devlin}, \binits{M.J.}},
\oauthor{\bsnm{Fissel}, \binits{L.M.}},
\oauthor{\bsnm{Fukui}, \binits{Y.}},
\oauthor{\bsnm{Galitzki}, \binits{N.}},
\oauthor{\bsnm{Gao}, \binits{J.}},
\oauthor{\bsnm{Groppi}, \binits{C.E.}},
\oauthor{\bsnm{Hillbrand}, \binits{S.}},
\oauthor{\bsnm{Hilton}, \binits{G.C.}},
\oauthor{\bsnm{Hubmayr}, \binits{J.}},
\oauthor{\bsnm{Irwin}, \binits{K.D.}},
\oauthor{\bsnm{Klein}, \binits{J.}},
\oauthor{\bsnm{Lanen}, \binits{J.V.}},
\oauthor{\bsnm{Li}, \binits{D.}},
\oauthor{\bsnm{Li}, \binits{Z.-Y.}},
\oauthor{\bsnm{Lourie}, \binits{N.P.}},
\oauthor{\bsnm{Mani}, \binits{H.}},
\oauthor{\bsnm{Martin}, \binits{P.G.}},
\oauthor{\bsnm{Mauskopf}, \binits{P.}},
\oauthor{\bsnm{Nakamura}, \binits{F.}},
\oauthor{\bsnm{Novak}, \binits{G.}},
\oauthor{\bsnm{Pappas}, \binits{D.P.}},
\oauthor{\bsnm{Pascale}, \binits{E.}},
\oauthor{\bsnm{Santos}, \binits{F.P.}},
\oauthor{\bsnm{Savini}, \binits{G.}},
\oauthor{\bsnm{Scott}, \binits{D.}},
\oauthor{\bsnm{Stanchfield}, \binits{S.}},
\oauthor{\bsnm{Ullom}, \binits{J.N.}},
\oauthor{\bsnm{Underhill}, \binits{M.}},
\oauthor{\bsnm{Vissers}, \binits{M.R.}},
\oauthor{\bsnm{Ward-Thompson}, \binits{D.}}:
The next-generation {BLASTPol} experiment.
In: Millimeter, Submillimeter, and Far-Infrared Detectors and Instrumentation for Astronomy {VII},
vol. 9153,
pp. 137--148.
{SPIE}.
\doiurl{10.1117/12.2054419} .
\url{https://www.spiedigitallibrary.org/conference-proceedings-of-spie/9153/91530H/The-next-generation-BLASTPol-experiment/10.1117/12.2054419.full}
\end{botherref}
\endbibitem

\bibitem[\protect\citeauthoryear{Adam et~al.}{}]{adam_nika2_2018}
\begin{botherref}
\oauthor{\bsnm{Adam}, \binits{R.}},
\oauthor{\bsnm{Adane}, \binits{A.}},
\oauthor{\bsnm{Ade}, \binits{P.A.R.}},
\oauthor{\bsnm{Andr\'e}, \binits{P.}},
\oauthor{\bsnm{Andrianasolo}, \binits{A.}},
\oauthor{\bsnm{Aussel}, \binits{H.}},
\oauthor{\bsnm{Beelen}, \binits{A.}},
\oauthor{\bsnm{Beno\^it}, \binits{A.}},
\oauthor{\bsnm{Bideaud}, \binits{A.}},
\oauthor{\bsnm{Billot}, \binits{N.}},
\oauthor{\bsnm{Bourrion}, \binits{O.}},
\oauthor{\bsnm{Bracco}, \binits{A.}},
\oauthor{\bsnm{Calvo}, \binits{M.}},
\oauthor{\bsnm{Catalano}, \binits{A.}},
\oauthor{\bsnm{Coiffard}, \binits{G.}},
\oauthor{\bsnm{Comis}, \binits{B.}},
\oauthor{\bsnm{Petris}, \binits{M.D.}},
\oauthor{\bsnm{D\'esert}, \binits{F.-X.}},
\oauthor{\bsnm{Doyle}, \binits{S.}},
\oauthor{\bsnm{Driessen}, \binits{E.F.C.}},
\oauthor{\bsnm{Evans}, \binits{R.}},
\oauthor{\bsnm{Goupy}, \binits{J.}},
\oauthor{\bsnm{Kramer}, \binits{C.}},
\oauthor{\bsnm{Lagache}, \binits{G.}},
\oauthor{\bsnm{Leclercq}, \binits{S.}},
\oauthor{\bsnm{Leggeri}, \binits{J.-P.}},
\oauthor{\bsnm{Lestrade}, \binits{J.-F.}},
\oauthor{\bsnm{Macías-Pérez}, \binits{J.F.}},
\oauthor{\bsnm{Mauskopf}, \binits{P.}},
\oauthor{\bsnm{Mayet}, \binits{F.}},
\oauthor{\bsnm{Maury}, \binits{A.}},
\oauthor{\bsnm{Monfardini}, \binits{A.}},
\oauthor{\bsnm{Navarro}, \binits{S.}},
\oauthor{\bsnm{Pascale}, \binits{E.}},
\oauthor{\bsnm{Perotto}, \binits{L.}},
\oauthor{\bsnm{Pisano}, \binits{G.}},
\oauthor{\bsnm{Ponthieu}, \binits{N.}},
\oauthor{\bsnm{Revéret}, \binits{V.}},
\oauthor{\bsnm{Rigby}, \binits{A.}},
\oauthor{\bsnm{Ritacco}, \binits{A.}},
\oauthor{\bsnm{Romero}, \binits{C.}},
\oauthor{\bsnm{Roussel}, \binits{H.}},
\oauthor{\bsnm{Ruppin}, \binits{F.}},
\oauthor{\bsnm{Schuster}, \binits{K.}},
\oauthor{\bsnm{Sievers}, \binits{A.}},
\oauthor{\bsnm{Triqueneaux}, \binits{S.}},
\oauthor{\bsnm{Tucker}, \binits{C.}},
\oauthor{\bsnm{Zylka}, \binits{R.}}:
The {NIKA}2 large-field-of-view millimetre continuum camera for the 30 m {IRAM} telescope
\textbf{609},
115
\doiurl{10.1051/0004-6361/201731503} .
Publisher: {EDP} Sciences
\end{botherref}
\endbibitem

\bibitem[\protect\citeauthoryear{Austermann et~al.}{}]{austermann_millimeter-wave_2018}
\begin{botherref}
\oauthor{\bsnm{Austermann}, \binits{J.E.}},
\oauthor{\bsnm{Beall}, \binits{J.A.}},
\oauthor{\bsnm{Bryan}, \binits{S.A.}},
\oauthor{\bsnm{Dober}, \binits{B.}},
\oauthor{\bsnm{Gao}, \binits{J.}},
\oauthor{\bsnm{Hilton}, \binits{G.}},
\oauthor{\bsnm{Hubmayr}, \binits{J.}},
\oauthor{\bsnm{Mauskopf}, \binits{P.}},
\oauthor{\bsnm{{McKenney}}, \binits{C.M.}},
\oauthor{\bsnm{Simon}, \binits{S.M.}},
\oauthor{\bsnm{Ullom}, \binits{J.N.}},
\oauthor{\bsnm{Vissers}, \binits{M.R.}},
\oauthor{\bsnm{Wilson}, \binits{G.W.}}:
Millimeter-wave polarimeters using kinetic inductance detectors for {TolTEC} and beyond
\textbf{193}(3),
10--10071090901819495
\doiurl{10.1007/s10909-018-1949-5}
\end{botherref}
\endbibitem

\bibitem[\protect\citeauthoryear{Wheeler et~al.}{}]{wheeler_broadband_2022}
\begin{botherref}
\oauthor{\bsnm{Wheeler}, \binits{J.}},
\oauthor{\bsnm{Austermann}, \binits{J.}},
\oauthor{\bsnm{Vissers}, \binits{M.}},
\oauthor{\bsnm{Beall}, \binits{J.}},
\oauthor{\bsnm{Gao}, \binits{J.}},
\oauthor{\bsnm{Imrek}, \binits{J.}},
\oauthor{\bsnm{Heilweil}, \binits{E.}},
\oauthor{\bsnm{Bennett}, \binits{D.}},
\oauthor{\bsnm{Gard}, \binits{J.}},
\oauthor{\bsnm{Lanen}, \binits{J.v.}},
\oauthor{\bsnm{Hubmayr}, \binits{J.}},
\oauthor{\bsnm{Ullom}, \binits{J.}}:
Broadband kinetic inductance detectors for far-{IR} observations.
In: Millimeter, Submillimeter, and Far-Infrared Detectors and Instrumentation for Astronomy {XI},
vol. 12190,
pp. 92--111.
{SPIE}.
\doiurl{10.1117/12.2630672} .
\url{https://www.spiedigitallibrary.org/conference-proceedings-of-spie/12190/1219006/Broadband-kinetic-inductance-detectors-for-far-IR-observations/10.1117/12.2630672.full}
\end{botherref}
\endbibitem

\bibitem[\protect\citeauthoryear{Mauskopf et~al.}{}]{mauskopf_photon-noise_2014}
\begin{botherref}
\oauthor{\bsnm{Mauskopf}, \binits{P.D.}},
\oauthor{\bsnm{Doyle}, \binits{S.}},
\oauthor{\bsnm{Barry}, \binits{P.}},
\oauthor{\bsnm{Rowe}, \binits{S.}},
\oauthor{\bsnm{Bidead}, \binits{A.}},
\oauthor{\bsnm{Ade}, \binits{P.A.R.}},
\oauthor{\bsnm{Tucker}, \binits{C.}},
\oauthor{\bsnm{Castillo}, \binits{E.}},
\oauthor{\bsnm{Monfardini}, \binits{A.}},
\oauthor{\bsnm{Goupy}, \binits{J.}},
\oauthor{\bsnm{Calvo}, \binits{M.}}:
Photon-noise limited performance in aluminum {LEKIDs}
\textbf{176}(3),
545--552
\doiurl{10.1007/s10909-013-1069-1}
\end{botherref}
\endbibitem

\bibitem[\protect\citeauthoryear{Hubmayr et~al.}{}]{hubmayr_photon-noise_2015}
\begin{botherref}
\oauthor{\bsnm{Hubmayr}, \binits{J.}},
\oauthor{\bsnm{Beall}, \binits{J.}},
\oauthor{\bsnm{Becker}, \binits{D.}},
\oauthor{\bsnm{Cho}, \binits{H.-M.}},
\oauthor{\bsnm{Devlin}, \binits{M.}},
\oauthor{\bsnm{Dober}, \binits{B.}},
\oauthor{\bsnm{Groppi}, \binits{C.}},
\oauthor{\bsnm{Hilton}, \binits{G.C.}},
\oauthor{\bsnm{Irwin}, \binits{K.D.}},
\oauthor{\bsnm{Li}, \binits{D.}},
\oauthor{\bsnm{Mauskopf}, \binits{P.}},
\oauthor{\bsnm{Pappas}, \binits{D.P.}},
\oauthor{\bsnm{Van~Lanen}, \binits{J.}},
\oauthor{\bsnm{Vissers}, \binits{M.R.}},
\oauthor{\bsnm{Wang}, \binits{Y.}},
\oauthor{\bsnm{Wei}, \binits{L.F.}},
\oauthor{\bsnm{Gao}, \binits{J.}}:
Photon-noise limited sensitivity in titanium nitride kinetic inductance detectors
\textbf{106}(7),
073505
\doiurl{10.1063/1.4913418}
\end{botherref}
\endbibitem

\bibitem[\protect\citeauthoryear{Monfardini et~al.}{}]{monfardini_dual-band_2011}
\begin{botherref}
\oauthor{\bsnm{Monfardini}, \binits{A.}},
\oauthor{\bsnm{Benoit}, \binits{A.}},
\oauthor{\bsnm{Bideaud}, \binits{A.}},
\oauthor{\bsnm{Swenson}, \binits{L.}},
\oauthor{\bsnm{Cruciani}, \binits{A.}},
\oauthor{\bsnm{Camus}, \binits{P.}},
\oauthor{\bsnm{Hoffmann}, \binits{C.}},
\oauthor{\bsnm{Désert}, \binits{F.X.}},
\oauthor{\bsnm{Doyle}, \binits{S.}},
\oauthor{\bsnm{Ade}, \binits{P.}},
\oauthor{\bsnm{Mauskopf}, \binits{P.}},
\oauthor{\bsnm{Tucker}, \binits{C.}},
\oauthor{\bsnm{Roesch}, \binits{M.}},
\oauthor{\bsnm{Leclercq}, \binits{S.}},
\oauthor{\bsnm{Schuster}, \binits{K.F.}},
\oauthor{\bsnm{Endo}, \binits{A.}},
\oauthor{\bsnm{Baryshev}, \binits{A.}},
\oauthor{\bsnm{Baselmans}, \binits{J.J.A.}},
\oauthor{\bsnm{Ferrari}, \binits{L.}},
\oauthor{\bsnm{Yates}, \binits{S.J.C.}},
\oauthor{\bsnm{Bourrion}, \binits{O.}},
\oauthor{\bsnm{Macias-Perez}, \binits{J.}},
\oauthor{\bsnm{Vescovi}, \binits{C.}},
\oauthor{\bsnm{Calvo}, \binits{M.}},
\oauthor{\bsnm{Giordano}, \binits{C.}}:
A dual-band millimeter-wave kinetic inductance camera for the {IRAM} 30 m telescope
\textbf{194}(2),
24
\doiurl{10.1088/0067-0049/194/2/24} .
Publisher: The American Astronomical Society
\end{botherref}
\endbibitem

\bibitem[\protect\citeauthoryear{Paiella et~al.}{}]{paiella_kinetic_2019}
\begin{botherref}
\oauthor{\bsnm{Paiella}, \binits{A.}},
\oauthor{\bsnm{Battistelli}, \binits{E.S.}},
\oauthor{\bsnm{Castellano}, \binits{M.G.}},
\oauthor{\bsnm{Colantoni}, \binits{I.}},
\oauthor{\bsnm{Columbro}, \binits{F.}},
\oauthor{\bsnm{Coppolecchia}, \binits{A.}},
\oauthor{\bsnm{D’Alessandro}, \binits{G.}},
\oauthor{\bsnm{Bernardis}, \binits{P.d.}},
\oauthor{\bsnm{Gordon}, \binits{S.}},
\oauthor{\bsnm{Lamagna}, \binits{L.}},
\oauthor{\bsnm{Mani}, \binits{H.}},
\oauthor{\bsnm{Masi}, \binits{S.}},
\oauthor{\bsnm{Mauskopf}, \binits{P.}},
\oauthor{\bsnm{Pettinari}, \binits{G.}},
\oauthor{\bsnm{Piacentini}, \binits{F.}},
\oauthor{\bsnm{Presta}, \binits{G.}}:
Kinetic inductance detectors and readout electronics for the {OLIMPO} experiment
\textbf{1182}(1),
012005
\doiurl{10.1088/1742-6596/1182/1/012005} .
Publisher: {IOP} Publishing
\end{botherref}
\endbibitem

\bibitem[\protect\citeauthoryear{Brien et~al.}{}]{brien_muscat_2018}
\begin{botherref}
\oauthor{\bsnm{Brien}, \binits{T.L.R.}},
\oauthor{\bsnm{Ade}, \binits{P.A.R.}},
\oauthor{\bsnm{Barry}, \binits{P.S.}},
\oauthor{\bsnm{Castillo-Domìnguez}, \binits{E.}},
\oauthor{\bsnm{Ferrusca}, \binits{D.}},
\oauthor{\bsnm{Gascard}, \binits{T.}},
\oauthor{\bsnm{Gómez}, \binits{V.}},
\oauthor{\bsnm{Hargrave}, \binits{P.C.}},
\oauthor{\bsnm{Hornsby}, \binits{A.L.}},
\oauthor{\bsnm{Hughes}, \binits{D.}},
\oauthor{\bsnm{Pascale}, \binits{E.}},
\oauthor{\bsnm{Parrianen}, \binits{J.D.A.}},
\oauthor{\bsnm{Perez}, \binits{A.}},
\oauthor{\bsnm{Rowe}, \binits{S.}},
\oauthor{\bsnm{Tucker}, \binits{C.}},
\oauthor{\bsnm{González}, \binits{S.V.}},
\oauthor{\bsnm{Doyle}, \binits{S.M.}}:
{MUSCAT}: the mexico-{UK} sub-millimetre camera for {AsTronomy}.
In: Millimeter, Submillimeter, and Far-Infrared Detectors and Instrumentation for Astronomy {IX},
vol. 10708,
pp. 173--181.
{SPIE}.
\doiurl{10.1117/12.2313697} .
\url{https://www.spiedigitallibrary.org/conference-proceedings-of-spie/10708/107080M/MUSCAT-the-Mexico-UK-Sub-Millimetre-Camera-for-AsTronomy/10.1117/12.2313697.full}
\end{botherref}
\endbibitem

\bibitem[\protect\citeauthoryear{Swenson et~al.}{}]{swenson_mako_2012}
\begin{botherref}
\oauthor{\bsnm{Swenson}, \binits{L.J.}},
\oauthor{\bsnm{Day}, \binits{P.K.}},
\oauthor{\bsnm{Dowell}, \binits{C.D.}},
\oauthor{\bsnm{Eom}, \binits{B.H.}},
\oauthor{\bsnm{Hollister}, \binits{M.I.}},
\oauthor{\bsnm{Jarnot}, \binits{R.}},
\oauthor{\bsnm{Kovács}, \binits{A.}},
\oauthor{\bsnm{Leduc}, \binits{H.G.}},
\oauthor{\bsnm{{McKenney}}, \binits{C.M.}},
\oauthor{\bsnm{Monroe}, \binits{R.}},
\oauthor{\bsnm{Mroczkowski}, \binits{T.}},
\oauthor{\bsnm{Nguyen}, \binits{H.T.}},
\oauthor{\bsnm{Zmuidzinas}, \binits{J.}}:
{MAKO}: a pathfinder instrument for on-sky demonstration of low-cost 350 micron imaging arrays.
In: Millimeter, Submillimeter, and Far-Infrared Detectors and Instrumentation for Astronomy {VI},
vol. 8452,
pp. 190--199.
{SPIE}.
\doiurl{10.1117/12.926223} .
\url{https://www.spiedigitallibrary.org/conference-proceedings-of-spie/8452/84520P/MAKO--a-pathfinder-instrument-for-on-sky-demonstration-of/10.1117/12.926223.full}
\end{botherref}
\endbibitem

\bibitem[\protect\citeauthoryear{Wilson et~al.}{}]{wilson_toltec_2020}
\begin{botherref}
\oauthor{\bsnm{Wilson}, \binits{G.W.}},
\oauthor{\bsnm{Abi-Saad}, \binits{S.}},
\oauthor{\bsnm{Ade}, \binits{P.}},
\oauthor{\bsnm{Aretxaga}, \binits{I.}},
\oauthor{\bsnm{Austermann}, \binits{J.}},
\oauthor{\bsnm{Ban}, \binits{Y.}},
\oauthor{\bsnm{Bardin}, \binits{J.}},
\oauthor{\bsnm{Beall}, \binits{J.}},
\oauthor{\bsnm{Berthoud}, \binits{M.}},
\oauthor{\bsnm{Bryan}, \binits{S.}},
\oauthor{\bsnm{Bussan}, \binits{J.}},
\oauthor{\bsnm{Castillo}, \binits{E.}},
\oauthor{\bsnm{Chavez}, \binits{M.}},
\oauthor{\bsnm{Contente}, \binits{R.}},
\oauthor{\bsnm{{DeNigris}}, \binits{N.S.}},
\oauthor{\bsnm{Dober}, \binits{B.}},
\oauthor{\bsnm{Eiben}, \binits{M.}},
\oauthor{\bsnm{Ferrusca}, \binits{D.}},
\oauthor{\bsnm{Fissel}, \binits{L.}},
\oauthor{\bsnm{Gao}, \binits{J.}},
\oauthor{\bsnm{Golec}, \binits{J.E.}},
\oauthor{\bsnm{Golina}, \binits{R.}},
\oauthor{\bsnm{Gomez}, \binits{A.}},
\oauthor{\bsnm{Gordon}, \binits{S.}},
\oauthor{\bsnm{Gutermuth}, \binits{R.}},
\oauthor{\bsnm{Hilton}, \binits{G.}},
\oauthor{\bsnm{Hosseini}, \binits{M.}},
\oauthor{\bsnm{Hubmayr}, \binits{J.}},
\oauthor{\bsnm{Hughes}, \binits{D.}},
\oauthor{\bsnm{Kuczarski}, \binits{S.}},
\oauthor{\bsnm{Lee}, \binits{D.}},
\oauthor{\bsnm{Lunde}, \binits{E.}},
\oauthor{\bsnm{Ma}, \binits{Z.}},
\oauthor{\bsnm{Mani}, \binits{H.}},
\oauthor{\bsnm{Mauskopf}, \binits{P.}},
\oauthor{\bsnm{{McCrackan}}, \binits{M.}},
\oauthor{\bsnm{{McKenney}}, \binits{C.}},
\oauthor{\bsnm{{McMahon}}, \binits{J.}},
\oauthor{\bsnm{Novak}, \binits{G.}},
\oauthor{\bsnm{Pisano}, \binits{G.}},
\oauthor{\bsnm{Pope}, \binits{A.}},
\oauthor{\bsnm{Ralston}, \binits{A.}},
\oauthor{\bsnm{Rodriguez}, \binits{I.}},
\oauthor{\bsnm{Sánchez-Argüelles}, \binits{D.}},
\oauthor{\bsnm{Schloerb}, \binits{F.P.}},
\oauthor{\bsnm{Simon}, \binits{S.}},
\oauthor{\bsnm{Sinclair}, \binits{A.}},
\oauthor{\bsnm{Souccar}, \binits{K.}},
\oauthor{\bsnm{Campos}, \binits{A.T.}},
\oauthor{\bsnm{Tucker}, \binits{C.}},
\oauthor{\bsnm{Ullom}, \binits{J.}},
\oauthor{\bsnm{Camp}, \binits{E.V.}},
\oauthor{\bsnm{Lanen}, \binits{J.V.}},
\oauthor{\bsnm{Velazquez}, \binits{M.}},
\oauthor{\bsnm{Vissers}, \binits{M.}},
\oauthor{\bsnm{Weeks}, \binits{E.}},
\oauthor{\bsnm{Yun}, \binits{M.S.}}:
The {TolTEC} camera: an overview of the instrument and in-lab testing results.
In: Millimeter, Submillimeter, and Far-Infrared Detectors and Instrumentation for Astronomy X,
vol. 11453,
p. 1145302.
{SPIE}.
\doiurl{10.1117/12.2562331} .
\url{https://www.spiedigitallibrary.org/conference-proceedings-of-spie/11453/1145302/The-TolTEC-camera--an-overview-of-the-instrument-and/10.1117/12.2562331.full}
\end{botherref}
\endbibitem

\bibitem[\protect\citeauthoryear{Mazin et~al.}{}]{mazin_arcons_2013}
\begin{botherref}
\oauthor{\bsnm{Mazin}, \binits{B.A.}},
\oauthor{\bsnm{Meeker}, \binits{S.R.}},
\oauthor{\bsnm{Strader}, \binits{M.J.}},
\oauthor{\bsnm{Szypryt}, \binits{P.}},
\oauthor{\bsnm{Marsden}, \binits{D.}},
\oauthor{\bsnm{Eyken}, \binits{J.C.v.}},
\oauthor{\bsnm{Duggan}, \binits{G.E.}},
\oauthor{\bsnm{Walter}, \binits{A.B.}},
\oauthor{\bsnm{Ulbricht}, \binits{G.}},
\oauthor{\bsnm{Johnson}, \binits{M.}},
\oauthor{\bsnm{Bumble}, \binits{B.}},
\oauthor{\bsnm{O’Brien}, \binits{K.}},
\oauthor{\bsnm{Stoughton}, \binits{C.}}:
{ARCONS}: A 2024 pixel optical through near-{IR} cryogenic imaging spectrophotometer
\textbf{125}(933),
1348
\doiurl{10.1086/674013} .
Publisher: {IOP} Publishing
\end{botherref}
\endbibitem

\bibitem[\protect\citeauthoryear{Parshley et~al.}{}]{parshley_ccat-prime_2018}
\begin{botherref}
\oauthor{\bsnm{Parshley}, \binits{S.C.}},
\oauthor{\bsnm{Kronshage}, \binits{J.}},
\oauthor{\bsnm{Blair}, \binits{J.}},
\oauthor{\bsnm{Herter}, \binits{T.}},
\oauthor{\bsnm{Nolta}, \binits{M.}},
\oauthor{\bsnm{Stacey}, \binits{G.J.}},
\oauthor{\bsnm{Bazarko}, \binits{A.}},
\oauthor{\bsnm{Bertoldi}, \binits{F.}},
\oauthor{\bsnm{Bustos}, \binits{R.}},
\oauthor{\bsnm{Campbell}, \binits{D.B.}},
\oauthor{\bsnm{Chapman}, \binits{S.}},
\oauthor{\bsnm{Cothard}, \binits{N.}},
\oauthor{\bsnm{Devlin}, \binits{M.}},
\oauthor{\bsnm{Erler}, \binits{J.}},
\oauthor{\bsnm{Fich}, \binits{M.}},
\oauthor{\bsnm{Gallardo}, \binits{P.A.}},
\oauthor{\bsnm{Giovanelli}, \binits{R.}},
\oauthor{\bsnm{Graf}, \binits{U.}},
\oauthor{\bsnm{Gramke}, \binits{S.}},
\oauthor{\bsnm{Haynes}, \binits{M.P.}},
\oauthor{\bsnm{Hills}, \binits{R.}},
\oauthor{\bsnm{Limon}, \binits{M.}},
\oauthor{\bsnm{Mangum}, \binits{J.G.}},
\oauthor{\bsnm{{McMahon}}, \binits{J.}},
\oauthor{\bsnm{Niemack}, \binits{M.D.}},
\oauthor{\bsnm{Nikola}, \binits{T.}},
\oauthor{\bsnm{Omlor}, \binits{M.}},
\oauthor{\bsnm{Riechers}, \binits{D.A.}},
\oauthor{\bsnm{Steeger}, \binits{K.}},
\oauthor{\bsnm{Stutzki}, \binits{J.}},
\oauthor{\bsnm{Vavagiakis}, \binits{E.M.}}:
{CCAT}-prime: a novel telescope for sub-millimeter astronomy.
In: Ground-based and Airborne Telescopes {VII},
vol. 10700,
pp. 1744--1758.
{SPIE}.
\doiurl{10.1117/12.2314046} .
\url{https://www.spiedigitallibrary.org/conference-proceedings-of-spie/10700/107005X/CCAT-prime-a-novel-telescope-for-sub-millimeter-astronomy/10.1117/12.2314046.full}
\end{botherref}
\endbibitem

\bibitem[\protect\citeauthoryear{Niemack}{}]{niemack_designs_2016}
\begin{botherref}
\oauthor{\bsnm{Niemack}, \binits{M.D.}}:
Designs for a large-aperture telescope to map the {CMB} 10× faster
\textbf{55}(7),
1688--1696
\doiurl{10.1364/AO.55.001688} .
Publisher: Optica Publishing Group
\end{botherref}
\endbibitem

\bibitem[\protect\citeauthoryear{Vavagiakis et~al.}{}]{vavagiakis_prime-cam_2018}
\begin{botherref}
\oauthor{\bsnm{Vavagiakis}, \binits{E.M.}},
\oauthor{\bsnm{Ahmed}, \binits{Z.}},
\oauthor{\bsnm{Ali}, \binits{A.}},
\oauthor{\bsnm{Basu}, \binits{K.}},
\oauthor{\bsnm{Battaglia}, \binits{N.}},
\oauthor{\bsnm{Bertoldi}, \binits{F.}},
\oauthor{\bsnm{Bond}, \binits{R.}},
\oauthor{\bsnm{Bustos}, \binits{R.}},
\oauthor{\bsnm{Chapman}, \binits{S.C.}},
\oauthor{\bsnm{Chung}, \binits{D.}},
\oauthor{\bsnm{Coppi}, \binits{G.}},
\oauthor{\bsnm{Cothard}, \binits{N.F.}},
\oauthor{\bsnm{Dicker}, \binits{S.}},
\oauthor{\bsnm{Duell}, \binits{C.J.}},
\oauthor{\bsnm{Duff}, \binits{S.M.}},
\oauthor{\bsnm{Erler}, \binits{J.}},
\oauthor{\bsnm{Fich}, \binits{M.}},
\oauthor{\bsnm{Galitzki}, \binits{N.}},
\oauthor{\bsnm{Gallardo}, \binits{P.A.}},
\oauthor{\bsnm{Henderson}, \binits{S.W.}},
\oauthor{\bsnm{Herter}, \binits{T.L.}},
\oauthor{\bsnm{Hilton}, \binits{G.}},
\oauthor{\bsnm{Hubmayr}, \binits{J.}},
\oauthor{\bsnm{Irwin}, \binits{K.D.}},
\oauthor{\bsnm{Koopman}, \binits{B.J.}},
\oauthor{\bsnm{{McMahon}}, \binits{J.}},
\oauthor{\bsnm{Murray}, \binits{N.}},
\oauthor{\bsnm{Niemack}, \binits{M.D.}},
\oauthor{\bsnm{Nikola}, \binits{T.}},
\oauthor{\bsnm{Nolta}, \binits{M.}},
\oauthor{\bsnm{Orlowski-Scherer}, \binits{J.}},
\oauthor{\bsnm{Parshley}, \binits{S.C.}},
\oauthor{\bsnm{Riechers}, \binits{D.A.}},
\oauthor{\bsnm{Rossi}, \binits{K.}},
\oauthor{\bsnm{Scott}, \binits{D.}},
\oauthor{\bsnm{Sierra}, \binits{C.}},
\oauthor{\bsnm{Silva-Feaver}, \binits{M.}},
\oauthor{\bsnm{Simon}, \binits{S.M.}},
\oauthor{\bsnm{Stacey}, \binits{G.J.}},
\oauthor{\bsnm{Stevens}, \binits{J.R.}},
\oauthor{\bsnm{Ullom}, \binits{J.N.}},
\oauthor{\bsnm{Vissers}, \binits{M.R.}},
\oauthor{\bsnm{Walker}, \binits{S.}},
\oauthor{\bsnm{Wollack}, \binits{E.J.}},
\oauthor{\bsnm{Xu}, \binits{Z.}},
\oauthor{\bsnm{Zhu}, \binits{N.}}:
Prime-cam: a first-light instrument for the {CCAT}-prime telescope.
In: Millimeter, Submillimeter, and Far-Infrared Detectors and Instrumentation for Astronomy {IX},
vol. 10708,
pp. 375--390.
{SPIE}.
\doiurl{10.1117/12.2313868} .
\url{https://www.spiedigitallibrary.org/conference-proceedings-of-spie/10708/107081U/Prime-Cam--a-first-light-instrument-for-the-CCAT/10.1117/12.2313868.full}
\end{botherref}
\endbibitem

\bibitem[\protect\citeauthoryear{Collaboration et~al.}{}]{collaboration_ccat-prime_2022}
\begin{botherref}
\oauthor{\bsnm{Collaboration}, \binits{C.-P.}},
\oauthor{\bsnm{Aravena}, \binits{M.}},
\oauthor{\bsnm{Austermann}, \binits{J.E.}},
\oauthor{\bsnm{Basu}, \binits{K.}},
\oauthor{\bsnm{Battaglia}, \binits{N.}},
\oauthor{\bsnm{Beringue}, \binits{B.}},
\oauthor{\bsnm{Bertoldi}, \binits{F.}},
\oauthor{\bsnm{Bigiel}, \binits{F.}},
\oauthor{\bsnm{Bond}, \binits{J.R.}},
\oauthor{\bsnm{Breysse}, \binits{P.C.}},
\oauthor{\bsnm{Broughton}, \binits{C.}},
\oauthor{\bsnm{Bustos}, \binits{R.}},
\oauthor{\bsnm{Chapman}, \binits{S.C.}},
\oauthor{\bsnm{Charmetant}, \binits{M.}},
\oauthor{\bsnm{Choi}, \binits{S.K.}},
\oauthor{\bsnm{Chung}, \binits{D.T.}},
\oauthor{\bsnm{Clark}, \binits{S.E.}},
\oauthor{\bsnm{Cothard}, \binits{N.F.}},
\oauthor{\bsnm{Crites}, \binits{A.T.}},
\oauthor{\bsnm{Dev}, \binits{A.}},
\oauthor{\bsnm{Douglas}, \binits{K.}},
\oauthor{\bsnm{Duell}, \binits{C.J.}},
\oauthor{\bsnm{Dünner}, \binits{R.}},
\oauthor{\bsnm{Ebina}, \binits{H.}},
\oauthor{\bsnm{Erler}, \binits{J.}},
\oauthor{\bsnm{Fich}, \binits{M.}},
\oauthor{\bsnm{Fissel}, \binits{L.M.}},
\oauthor{\bsnm{Foreman}, \binits{S.}},
\oauthor{\bsnm{Freundt}, \binits{R.G.}},
\oauthor{\bsnm{Gallardo}, \binits{P.A.}},
\oauthor{\bsnm{Gao}, \binits{J.}},
\oauthor{\bsnm{García}, \binits{P.}},
\oauthor{\bsnm{Giovanelli}, \binits{R.}},
\oauthor{\bsnm{Golec}, \binits{J.E.}},
\oauthor{\bsnm{Groppi}, \binits{C.E.}},
\oauthor{\bsnm{Haynes}, \binits{M.P.}},
\oauthor{\bsnm{Henke}, \binits{D.}},
\oauthor{\bsnm{Hensley}, \binits{B.}},
\oauthor{\bsnm{Herter}, \binits{T.}},
\oauthor{\bsnm{Higgins}, \binits{R.}},
\oauthor{\bsnm{Hložek}, \binits{R.}},
\oauthor{\bsnm{Huber}, \binits{A.}},
\oauthor{\bsnm{Huber}, \binits{Z.}},
\oauthor{\bsnm{Hubmayr}, \binits{J.}},
\oauthor{\bsnm{Jackson}, \binits{R.}},
\oauthor{\bsnm{Johnstone}, \binits{D.}},
\oauthor{\bsnm{Karoumpis}, \binits{C.}},
\oauthor{\bsnm{Keating}, \binits{L.C.}},
\oauthor{\bsnm{Komatsu}, \binits{E.}},
\oauthor{\bsnm{Li}, \binits{Y.}},
\oauthor{\bsnm{Magnelli}, \binits{B.}},
\oauthor{\bsnm{Matthews}, \binits{B.C.}},
\oauthor{\bsnm{Mauskopf}, \binits{P.D.}},
\oauthor{\bsnm{{McMahon}}, \binits{J.J.}},
\oauthor{\bsnm{Meerburg}, \binits{P.D.}},
\oauthor{\bsnm{Meyers}, \binits{J.}},
\oauthor{\bsnm{Muralidhara}, \binits{V.}},
\oauthor{\bsnm{Murray}, \binits{N.W.}},
\oauthor{\bsnm{Niemack}, \binits{M.D.}},
\oauthor{\bsnm{Nikola}, \binits{T.}},
\oauthor{\bsnm{Okada}, \binits{Y.}},
\oauthor{\bsnm{Puddu}, \binits{R.}},
\oauthor{\bsnm{Riechers}, \binits{D.A.}},
\oauthor{\bsnm{Rosolowsky}, \binits{E.}},
\oauthor{\bsnm{Rossi}, \binits{K.}},
\oauthor{\bsnm{Rotermund}, \binits{K.}},
\oauthor{\bsnm{Roy}, \binits{A.}},
\oauthor{\bsnm{Sadavoy}, \binits{S.I.}},
\oauthor{\bsnm{Schaaf}, \binits{R.}},
\oauthor{\bsnm{Schilke}, \binits{P.}},
\oauthor{\bsnm{Scott}, \binits{D.}},
\oauthor{\bsnm{Simon}, \binits{R.}},
\oauthor{\bsnm{Sinclair}, \binits{A.K.}},
\oauthor{\bsnm{Sivakoff}, \binits{G.R.}},
\oauthor{\bsnm{Stacey}, \binits{G.J.}},
\oauthor{\bsnm{Stutz}, \binits{A.M.}},
\oauthor{\bsnm{Stutzki}, \binits{J.}},
\oauthor{\bsnm{Tahani}, \binits{M.}},
\oauthor{\bsnm{Thanjavur}, \binits{K.}},
\oauthor{\bsnm{Timmermann}, \binits{R.A.}},
\oauthor{\bsnm{Ullom}, \binits{J.N.}},
\oauthor{\bsnm{Engelen}, \binits{A.v.}},
\oauthor{\bsnm{Vavagiakis}, \binits{E.M.}},
\oauthor{\bsnm{Vissers}, \binits{M.R.}},
\oauthor{\bsnm{Wheeler}, \binits{J.D.}},
\oauthor{\bsnm{White}, \binits{S.D.M.}},
\oauthor{\bsnm{Zhu}, \binits{Y.}},
\oauthor{\bsnm{Zou}, \binits{B.}}:
{CCAT}-prime collaboration: Science goals and forecasts with {P}rime-{C}am on the {F}red {Y}oung {S}ubmillimeter {T}elescope
\textbf{264}(1),
7
\doiurl{10.3847/1538-4365/ac9838} .
Publisher: The American Astronomical Society
\end{botherref}
\endbibitem

\bibitem[\protect\citeauthoryear{Choi et~al.}{}]{choi_sensitivity_2020}
\begin{botherref}
\oauthor{\bsnm{Choi}, \binits{S.K.}},
\oauthor{\bsnm{Austermann}, \binits{J.}},
\oauthor{\bsnm{Basu}, \binits{K.}},
\oauthor{\bsnm{Battaglia}, \binits{N.}},
\oauthor{\bsnm{Bertoldi}, \binits{F.}},
\oauthor{\bsnm{Chung}, \binits{D.T.}},
\oauthor{\bsnm{Cothard}, \binits{N.F.}},
\oauthor{\bsnm{Duff}, \binits{S.}},
\oauthor{\bsnm{Duell}, \binits{C.J.}},
\oauthor{\bsnm{Gallardo}, \binits{P.A.}},
\oauthor{\bsnm{Gao}, \binits{J.}},
\oauthor{\bsnm{Herter}, \binits{T.}},
\oauthor{\bsnm{Hubmayr}, \binits{J.}},
\oauthor{\bsnm{Niemack}, \binits{M.D.}},
\oauthor{\bsnm{Nikola}, \binits{T.}},
\oauthor{\bsnm{Riechers}, \binits{D.}},
\oauthor{\bsnm{Rossi}, \binits{K.}},
\oauthor{\bsnm{Stacey}, \binits{G.J.}},
\oauthor{\bsnm{Stevens}, \binits{J.R.}},
\oauthor{\bsnm{Vavagiakis}, \binits{E.M.}},
\oauthor{\bsnm{Vissers}, \binits{M.}},
\oauthor{\bsnm{Walker}, \binits{S.}}:
Sensitivity of the {P}rime-{C}am instrument on the {CCAT}-prime telescope
\textbf{199}(3),
1089--1097
\doiurl{10.1007/s10909-020-02428-z}
\end{botherref}
\endbibitem

\bibitem[\protect\citeauthoryear{Vavagiakis et~al.}{}]{vavagiakis_ccat-prime_2022}
\begin{botherref}
\oauthor{\bsnm{Vavagiakis}, \binits{E.M.}},
\oauthor{\bsnm{Duell}, \binits{C.J.}},
\oauthor{\bsnm{Austermann}, \binits{J.}},
\oauthor{\bsnm{Beall}, \binits{J.}},
\oauthor{\bsnm{Bhandarkar}, \binits{T.}},
\oauthor{\bsnm{Chapman}, \binits{S.C.}},
\oauthor{\bsnm{Choi}, \binits{S.K.}},
\oauthor{\bsnm{Coppi}, \binits{G.}},
\oauthor{\bsnm{Dicker}, \binits{S.}},
\oauthor{\bsnm{Devlin}, \binits{M.}},
\oauthor{\bsnm{Freundt}, \binits{R.G.}},
\oauthor{\bsnm{Gao}, \binits{J.}},
\oauthor{\bsnm{Groppi}, \binits{C.}},
\oauthor{\bsnm{Herter}, \binits{T.L.}},
\oauthor{\bsnm{Huber}, \binits{Z.B.}},
\oauthor{\bsnm{Hubmayr}, \binits{J.}},
\oauthor{\bsnm{Johnstone}, \binits{D.}},
\oauthor{\bsnm{Keller}, \binits{B.}},
\oauthor{\bsnm{Kofman}, \binits{A.M.}},
\oauthor{\bsnm{Li}, \binits{Y.}},
\oauthor{\bsnm{Mauskopf}, \binits{P.}},
\oauthor{\bsnm{{McMahon}}, \binits{J.}},
\oauthor{\bsnm{Moore}, \binits{J.}},
\oauthor{\bsnm{Murphy}, \binits{C.C.}},
\oauthor{\bsnm{Niemack}, \binits{M.D.}},
\oauthor{\bsnm{Nikola}, \binits{T.}},
\oauthor{\bsnm{Orlowski-Scherer}, \binits{J.}},
\oauthor{\bsnm{Rossi}, \binits{K.M.}},
\oauthor{\bsnm{Sinclair}, \binits{A.K.}},
\oauthor{\bsnm{Stacey}, \binits{G.J.}},
\oauthor{\bsnm{Ullom}, \binits{J.}},
\oauthor{\bsnm{Vissers}, \binits{M.}},
\oauthor{\bsnm{Wheeler}, \binits{J.}},
\oauthor{\bsnm{Xu}, \binits{Z.}},
\oauthor{\bsnm{Zhu}, \binits{N.}},
\oauthor{\bsnm{Zou}, \binits{B.}}:
{CCAT}-prime: design of the mod-cam receiver and 280 {GHz} {MKID} instrument module.
In: Millimeter, Submillimeter, and Far-Infrared Detectors and Instrumentation for Astronomy {XI},
vol. 12190,
pp. 61--76.
{SPIE}.
\doiurl{10.1117/12.2630115} .
\url{https://www.spiedigitallibrary.org/conference-proceedings-of-spie/12190/1219004/CCAT-prime--design-of-the-Mod-Cam-receiver-and/10.1117/12.2630115.full}
\end{botherref}
\endbibitem

\bibitem[\protect\citeauthoryear{Duell et~al.}{}]{duell_ccat-prime_2020}
\begin{botherref}
\oauthor{\bsnm{Duell}, \binits{C.J.}},
\oauthor{\bsnm{Vavagiakis}, \binits{E.M.}},
\oauthor{\bsnm{Austermann}, \binits{J.}},
\oauthor{\bsnm{Chapman}, \binits{S.C.}},
\oauthor{\bsnm{Choi}, \binits{S.K.}},
\oauthor{\bsnm{Cothard}, \binits{N.F.}},
\oauthor{\bsnm{Dober}, \binits{B.}},
\oauthor{\bsnm{Gallardo}, \binits{P.}},
\oauthor{\bsnm{Gao}, \binits{J.}},
\oauthor{\bsnm{Groppi}, \binits{C.}},
\oauthor{\bsnm{Herter}, \binits{T.L.}},
\oauthor{\bsnm{Stacey}, \binits{G.J.}},
\oauthor{\bsnm{Huber}, \binits{Z.}},
\oauthor{\bsnm{Hubmayr}, \binits{J.}},
\oauthor{\bsnm{Johnstone}, \binits{D.}},
\oauthor{\bsnm{Li}, \binits{Y.}},
\oauthor{\bsnm{Mauskopf}, \binits{P.}},
\oauthor{\bsnm{{McMahon}}, \binits{J.}},
\oauthor{\bsnm{Niemack}, \binits{M.D.}},
\oauthor{\bsnm{Nikola}, \binits{T.}},
\oauthor{\bsnm{Rossi}, \binits{K.}},
\oauthor{\bsnm{Simon}, \binits{S.}},
\oauthor{\bsnm{Sinclair}, \binits{A.K.}},
\oauthor{\bsnm{Vissers}, \binits{M.}},
\oauthor{\bsnm{Wheeler}, \binits{J.}},
\oauthor{\bsnm{Zou}, \binits{B.}}:
{CCAT}-prime: Designs and status of the first light 280 {GHz} {MKID} array and mod-cam receiver.
In: Millimeter, Submillimeter, and Far-Infrared Detectors and Instrumentation for Astronomy X,
vol. 11453,
pp. 235--243.
{SPIE}.
\doiurl{10.1117/12.2562757} .
\url{https://www.spiedigitallibrary.org/conference-proceedings-of-spie/11453/114531F/CCAT-prime--Designs-and-status-of-the-first-light/10.1117/12.2562757.full}
\end{botherref}
\endbibitem

\bibitem[\protect\citeauthoryear{Austermann et~al.}{}]{austermann_toltec_2020}
\begin{botherref}
\oauthor{\bsnm{Austermann}, \binits{J.E.}},
\oauthor{\bsnm{Ban}, \binits{Y.}},
\oauthor{\bsnm{Beall}, \binits{J.}},
\oauthor{\bsnm{Berthoud}, \binits{M.}},
\oauthor{\bsnm{Contente}, \binits{R.}},
\oauthor{\bsnm{{DeNigris}}, \binits{N.}},
\oauthor{\bsnm{Dober}, \binits{B.}},
\oauthor{\bsnm{Gao}, \binits{J.}},
\oauthor{\bsnm{Hilton}, \binits{G.}},
\oauthor{\bsnm{Hubmayr}, \binits{J.}},
\oauthor{\bsnm{Kuczarski}, \binits{S.}},
\oauthor{\bsnm{Lee}, \binits{D.}},
\oauthor{\bsnm{Lunde}, \binits{E.}},
\oauthor{\bsnm{Ma}, \binits{Z.}},
\oauthor{\bsnm{Mauskopf}, \binits{P.}},
\oauthor{\bsnm{{McCrackan}}, \binits{M.}},
\oauthor{\bsnm{{McKenney}}, \binits{C.}},
\oauthor{\bsnm{{McMahon}}, \binits{J.}},
\oauthor{\bsnm{Novak}, \binits{G.}},
\oauthor{\bsnm{Abi-Saad}, \binits{S.}},
\oauthor{\bsnm{Simon}, \binits{S.}},
\oauthor{\bsnm{Souccar}, \binits{K.}},
\oauthor{\bsnm{Ullom}, \binits{J.}},
\oauthor{\bsnm{Camp}, \binits{E.V.}},
\oauthor{\bsnm{Lanen}, \binits{J.V.}},
\oauthor{\bsnm{Vissers}, \binits{M.}},
\oauthor{\bsnm{Wilson}, \binits{G.}}:
{TolTEC} focal plane arrays: design, characterization, and performance of kilopixel {MKID} focal planes.
In: Millimeter, Submillimeter, and Far-Infrared Detectors and Instrumentation for Astronomy X,
vol. 11453,
p. 114530.
{SPIE}.
\doiurl{10.1117/12.2562086} .
\url{https://www.spiedigitallibrary.org/conference-proceedings-of-spie/11453/114530A/TolTEC-focal-plane-arrays--design-characterization-and-performance-of/10.1117/12.2562086.full}
\end{botherref}
\endbibitem

\bibitem[\protect\citeauthoryear{Simon et~al.}{}]{simon_feedhorn_2018}
\begin{botherref}
\oauthor{\bsnm{Simon}, \binits{S.M.}},
\oauthor{\bsnm{Golec}, \binits{J.E.}},
\oauthor{\bsnm{Ali}, \binits{A.}},
\oauthor{\bsnm{Austermann}, \binits{J.}},
\oauthor{\bsnm{Beall}, \binits{J.A.}},
\oauthor{\bsnm{Bruno}, \binits{S.M.M.}},
\oauthor{\bsnm{Choi}, \binits{S.K.}},
\oauthor{\bsnm{Crowley}, \binits{K.T.}},
\oauthor{\bsnm{Dicker}, \binits{S.}},
\oauthor{\bsnm{Dober}, \binits{B.}},
\oauthor{\bsnm{Duff}, \binits{S.M.}},
\oauthor{\bsnm{Healy}, \binits{E.}},
\oauthor{\bsnm{Hill}, \binits{C.A.}},
\oauthor{\bsnm{Ho}, \binits{S.-P.P.}},
\oauthor{\bsnm{Hubmayr}, \binits{J.}},
\oauthor{\bsnm{Li}, \binits{Y.}},
\oauthor{\bsnm{Lungu}, \binits{M.}},
\oauthor{\bsnm{{McMahon}}, \binits{J.}},
\oauthor{\bsnm{Orlowski-Scherer}, \binits{J.}},
\oauthor{\bsnm{Salatino}, \binits{M.}},
\oauthor{\bsnm{Staggs}, \binits{S.}},
\oauthor{\bsnm{Wollack}, \binits{E.J.}},
\oauthor{\bsnm{Xu}, \binits{Z.}},
\oauthor{\bsnm{Zhu}, \binits{N.}}:
Feedhorn development and scalability for simons observatory and beyond.
In: Millimeter, Submillimeter, and Far-Infrared Detectors and Instrumentation for Astronomy {IX},
vol. 10708,
pp. 1145--1156.
{SPIE}.
\doiurl{10.1117/12.2313405} .
\url{https://www.spiedigitallibrary.org/conference-proceedings-of-spie/10708/107084B/Feedhorn-development-and-scalability-for-Simons-Observatory-and-beyond/10.1117/12.2313405.full}
\end{botherref}
\endbibitem

\bibitem[\protect\citeauthoryear{Choi et~al.}{}]{choi_ccat-prime_2022}
\begin{botherref}
\oauthor{\bsnm{Choi}, \binits{S.K.}},
\oauthor{\bsnm{Duell}, \binits{C.J.}},
\oauthor{\bsnm{Austermann}, \binits{J.}},
\oauthor{\bsnm{Cothard}, \binits{N.F.}},
\oauthor{\bsnm{Gao}, \binits{J.}},
\oauthor{\bsnm{Freundt}, \binits{R.G.}},
\oauthor{\bsnm{Groppi}, \binits{C.}},
\oauthor{\bsnm{Herter}, \binits{T.}},
\oauthor{\bsnm{Hubmayr}, \binits{J.}},
\oauthor{\bsnm{Huber}, \binits{Z.B.}},
\oauthor{\bsnm{Keller}, \binits{B.}},
\oauthor{\bsnm{Li}, \binits{Y.}},
\oauthor{\bsnm{Mauskopf}, \binits{P.}},
\oauthor{\bsnm{Niemack}, \binits{M.D.}},
\oauthor{\bsnm{Nikola}, \binits{T.}},
\oauthor{\bsnm{Rossi}, \binits{K.}},
\oauthor{\bsnm{Sinclair}, \binits{A.}},
\oauthor{\bsnm{Stacey}, \binits{G.J.}},
\oauthor{\bsnm{Vavagiakis}, \binits{E.M.}},
\oauthor{\bsnm{Vissers}, \binits{M.}},
\oauthor{\bsnm{Tucker}, \binits{C.}},
\oauthor{\bsnm{Weeks}, \binits{E.}},
\oauthor{\bsnm{Wheeler}, \binits{J.}}:
{CCAT}-prime: Characterization of the first 280 {GHz} {MKID} array for prime-cam
\textbf{209}(5),
849--856
\doiurl{10.1007/s10909-022-02787-9}
\end{botherref}
\endbibitem

\bibitem[\protect\citeauthoryear{Austermann et~al.}{}]{austermann_aluminum-based_2022}
\begin{botherref}
\oauthor{\bsnm{Austermann}, \binits{J.E.}},
\oauthor{\bsnm{Beall}, \binits{J.}},
\oauthor{\bsnm{Gao}, \binits{J.}},
\oauthor{\bsnm{Hubmayr}, \binits{J.}},
\oauthor{\bsnm{Ullom}, \binits{J.}},
\oauthor{\bsnm{Vissers}, \binits{M.}},
\oauthor{\bsnm{Wheeler}, \binits{J.}}:
Aluminum-based millimeter-wave kinetic inductance detectors on 150{\textasciitilde}mm diameter substrates.
In: Millimeter, Submillimeter, and Far-Infrared Detectors and Instrumentation for Astronomy {XI},
vol. {PC}12190,
p. 121900.
{SPIE}.
\doiurl{10.1117/12.2630512} .
\url{https://www.spiedigitallibrary.org/conference-proceedings-of-spie/PC12190/PC121900X/Aluminum-based-millimeter-wave-kinetic-inductance-detectors-on-150mm-diameter/10.1117/12.2630512.full}
\end{botherref}
\endbibitem

\bibitem[\protect\citeauthoryear{Liu et~al.}{}]{liu_superconducting_2017}
\begin{botherref}
\oauthor{\bsnm{Liu}, \binits{X.}},
\oauthor{\bsnm{Guo}, \binits{W.}},
\oauthor{\bsnm{Wang}, \binits{Y.}},
\oauthor{\bsnm{Dai}, \binits{M.}},
\oauthor{\bsnm{Wei}, \binits{L.F.}},
\oauthor{\bsnm{Dober}, \binits{B.}},
\oauthor{\bsnm{{McKenney}}, \binits{C.M.}},
\oauthor{\bsnm{Hilton}, \binits{G.C.}},
\oauthor{\bsnm{Hubmayr}, \binits{J.}},
\oauthor{\bsnm{Austermann}, \binits{J.E.}},
\oauthor{\bsnm{Ullom}, \binits{J.N.}},
\oauthor{\bsnm{Gao}, \binits{J.}},
\oauthor{\bsnm{Vissers}, \binits{M.R.}}:
Superconducting micro-resonator arrays with ideal frequency spacing
\textbf{111}(25),
252601
\doiurl{10.1063/1.5016190}
\end{botherref}
\endbibitem

\bibitem[\protect\citeauthoryear{Swenson et~al.}{}]{swenson_operation_2013}
\begin{botherref}
\oauthor{\bsnm{Swenson}, \binits{L.J.}},
\oauthor{\bsnm{Day}, \binits{P.K.}},
\oauthor{\bsnm{Eom}, \binits{B.H.}},
\oauthor{\bsnm{Leduc}, \binits{H.G.}},
\oauthor{\bsnm{Llombart}, \binits{N.}},
\oauthor{\bsnm{{McKenney}}, \binits{C.M.}},
\oauthor{\bsnm{Noroozian}, \binits{O.}},
\oauthor{\bsnm{Zmuidzinas}, \binits{J.}}:
Operation of a titanium nitride superconducting microresonator detector in the nonlinear regime
\textbf{113}(10),
104501
\doiurl{10.1063/1.4794808}
\end{botherref}
\endbibitem

\bibitem[\protect\citeauthoryear{Flanigan et~al.}{}]{flanigan_photon_2016}
\begin{botherref}
\oauthor{\bsnm{Flanigan}, \binits{D.}},
\oauthor{\bsnm{{McCarrick}}, \binits{H.}},
\oauthor{\bsnm{Jones}, \binits{G.}},
\oauthor{\bsnm{Johnson}, \binits{B.R.}},
\oauthor{\bsnm{Abitbol}, \binits{M.H.}},
\oauthor{\bsnm{Ade}, \binits{P.}},
\oauthor{\bsnm{Araujo}, \binits{D.}},
\oauthor{\bsnm{Bradford}, \binits{K.}},
\oauthor{\bsnm{Cantor}, \binits{R.}},
\oauthor{\bsnm{Che}, \binits{G.}},
\oauthor{\bsnm{Day}, \binits{P.}},
\oauthor{\bsnm{Doyle}, \binits{S.}},
\oauthor{\bsnm{Kjellstrand}, \binits{C.B.}},
\oauthor{\bsnm{Leduc}, \binits{H.}},
\oauthor{\bsnm{Limon}, \binits{M.}},
\oauthor{\bsnm{Luu}, \binits{V.}},
\oauthor{\bsnm{Mauskopf}, \binits{P.}},
\oauthor{\bsnm{Miller}, \binits{A.}},
\oauthor{\bsnm{Mroczkowski}, \binits{T.}},
\oauthor{\bsnm{Tucker}, \binits{C.}},
\oauthor{\bsnm{Zmuidzinas}, \binits{J.}}:
Photon noise from chaotic and coherent millimeter-wave sources measured with horn-coupled, aluminum lumped-element kinetic inductance detectors
\textbf{108}(8),
083504
\doiurl{10.1063/1.4942804}
\end{botherref}
\endbibitem

\bibitem[\protect\citeauthoryear{de~Visser et~al.}{}]{de_visser_evidence_2014}
\begin{botherref}
\oauthor{\bsnm{Visser}, \binits{P.J.}},
\oauthor{\bsnm{Goldie}, \binits{D.J.}},
\oauthor{\bsnm{Diener}, \binits{P.}},
\oauthor{\bsnm{Withington}, \binits{S.}},
\oauthor{\bsnm{Baselmans}, \binits{J.J.A.}},
\oauthor{\bsnm{Klapwijk}, \binits{T.M.}}:
Evidence of a nonequilibrium distribution of quasiparticles in the microwave response of a superconducting aluminum resonator
\textbf{112}(4),
047004
\doiurl{10.1103/PhysRevLett.112.047004} .
Publisher: American Physical Society
\end{botherref}
\endbibitem

\bibitem[\protect\citeauthoryear{Sinclair et~al.}{}]{sinclair_ccat-prime_2022}
\begin{botherref}
\oauthor{\bsnm{Sinclair}, \binits{A.K.}},
\oauthor{\bsnm{Stephenson}, \binits{R.C.}},
\oauthor{\bsnm{Roberson}, \binits{C.A.}},
\oauthor{\bsnm{Weeks}, \binits{E.L.}},
\oauthor{\bsnm{Burgoyne}, \binits{J.}},
\oauthor{\bsnm{Huber}, \binits{A.I.}},
\oauthor{\bsnm{Mauskopf}, \binits{P.M.}},
\oauthor{\bsnm{Chapman}, \binits{S.C.}},
\oauthor{\bsnm{Austermann}, \binits{J.E.}},
\oauthor{\bsnm{Choi}, \binits{S.K.}},
\oauthor{\bsnm{Duell}, \binits{C.J.}},
\oauthor{\bsnm{Fich}, \binits{M.}},
\oauthor{\bsnm{Groppi}, \binits{C.E.}},
\oauthor{\bsnm{Huber}, \binits{Z.}},
\oauthor{\bsnm{Niemack}, \binits{M.D.}},
\oauthor{\bsnm{Nikola}, \binits{T.}},
\oauthor{\bsnm{Rossi}, \binits{K.M.}},
\oauthor{\bsnm{Sriram}, \binits{A.}},
\oauthor{\bsnm{Stacey}, \binits{G.J.}},
\oauthor{\bsnm{Szakiel}, \binits{E.}},
\oauthor{\bsnm{Tsuchitori}, \binits{J.}},
\oauthor{\bsnm{Vavagiakis}, \binits{E.M.}},
\oauthor{\bsnm{Wheeler}, \binits{J.D.}}:
{CCAT}-prime: {RFSoC} based readout for frequency multiplexed kinetic inductance detectors.
In: Millimeter, Submillimeter, and Far-Infrared Detectors and Instrumentation for Astronomy {XI},
vol. 12190,
pp. 444--463.
{SPIE}.
\doiurl{10.1117/12.2629722} .
\url{https://www.spiedigitallibrary.org/conference-proceedings-of-spie/12190/121900W/CCAT-prime--RFSoC-based-readout-for-frequency-multiplexed-kinetic/10.1117/12.2629722.full}
\end{botherref}
\endbibitem

\end{thebibliography}

\end{document}